\documentstyle[prd,aps]{revtex}

\begin{document}
\draft

\title{Generalized Scale Invariant Gravity}
\author{Shih-Yuin Lin\footnote{Electronic address: 
{\tt sylin@phys.nctu.edu.tw}}}
\address{Department of Electro-Physics and Institute of Physics, Chiao-Tung 
University, Hsinchu, Taiwan}
\author{Kin-Wang Ng\footnote{Electronic address: {\tt nkw@phys.sinica.edu.tw}}}
\address{Institute of Physics, Academia Sinica, Taipei, Taiwan}
\date{August 1997}
\maketitle

\begin{abstract}
We generalize the scale invariant gravity by allowing a negative kinetic 
energy term for the classical scalar field. This gives birth to a new 
scalar-tensor theory of gravity, in which the scalar field is in fact
an auxiliary field. For a pure gravity theory without matter, 
the scale symmetric phase represents an equivalent 
class of gravity theories, which the Einstein gravity plus a cosmological 
constant belongs to under a special gauge choice. The one-loop quantum 
correction of the theory is calculated by using the Vilkovisky-DeWitt's 
method. We find that the scale symmetry is broken dynamically, and that 
the Einstein gravity is the ground state of the broken phase. We also 
briefly discuss the consequent cosmological implications. It is shown that
the time-delay experiment restricts the present universe to be very close
to the ground state.
\end{abstract}


\section{Introduction}

The classical Einstein gravity, dictated by the symmetry of general 
transformation of space-time coordinates based on the equivalence principle, 
agrees well with current observed data. However, it is not unreasonable to
ponder other alternatives or modified theories of gravity. Theoretically,
the coordinate-reparametrization symmetry is not garanteed to be the ultimate 
symmetry of gravity. For example, in contrast with the other three fundamental
forces, the quantum theory of Einstein gravity is non-renormalizable and the
coordinate-reparametrization symmetry is not enough to cancel all the 
divergences in quantum corrections. If we believe that physical theories
should be self-consistent, symmetries in addition to the reparametrization
symmetry should be involved in gravity.  

Inspired by the idea of the {\it relativity of motion} in Einstein's 
geometrical theory of space-time, Weyl\cite{weyl} introduced an additional
principle, the {\it relativity of magnitude}, which is realized by
the local scale symmetry of space-time. A bonus was that the
vector field presented in Weyl gravity looks like the electromagnetic
field in Maxwell's theory, so that all known interactions at that time
were hopeful to be unified in a geometrical sense.
Unfortunately, Weyl's hypothesis had been criticized by Einstein because it 
implies that the frequency of spectral lines emitted by atoms would not remain 
constant but would depend on their history. The reason behind 
is that quantum phenomena provide an absolute standard of scale.
An atomic clock measures time in an absolute way and an absolute standard 
of length is given by taking the light-speed to be unity. 
So, the scale symmetry seems not to be appreciated by the quantum nature.

This problem can be resolved by either introducing a scale symmetry breaking 
mechanism\cite{deser,smolin} or following Dirac's suggestion\cite{dirac}.
Dirac supposed that the Einstein equations refer to an space-time interval
connecting two neighbouring points which is not the same as the interval 
measured by atomic apparatus. In this paper we will follow the former wisdom. 
First of all, we will introduce the classical theory of scale invariant 
gravity. Then, we will present a generalized form of the theory, 
and show that it contains a new scalar-tensor theory of gravity. 
To implement the scale symmetry breaking, we will calculate the one-loop 
correction of the scalar-tensor action. It will be shown that the quantum 
effects would indeed break the scale symmetry dynamically. In order to avoid 
ambiguities arising from field reparametrizations or gauge transformations 
in doing the quantum corrections, we will adopt the Vilkovisky-DeWitt (VD)
\cite{vilkov,dewitt} method to handle the symmetries of general coordinate 
transformation as well as scale transformation. 
                                  
The paper is organized as follows. In Section II, we introduce the scale 
transformation and the classical scale invariant gravity. A new classical
theory of the scale invariant gravity is proposed in Section III. 
In Section IV, we calculate the Vilkovisky-DeWitt one-loop effective 
potential of the theory, with detailed calculations shown in Appendix.
In Section V, we briefly discuss the cosmological consequences.

\section{Classical Scale Invariant Gravity}

Let us begin with a covariant derivative in curved space-time,
\begin{equation}
  D_\mu V^\nu = \partial_\mu V^\nu - \Gamma_{\mu\rho}^\nu V^\rho,
\end{equation}
of a vector field $V_\mu$. As usual, $\Gamma_{\mu\rho}^\nu $ is the 
Christoffel symbol defined by
\begin{equation}
  \Gamma_{\mu\rho}^\nu \equiv {1\over 2}g^{\nu\sigma}\left( g_{\sigma\mu ,\rho}
    + g_{\sigma\rho ,\mu} - g_{\mu\rho,\sigma}\right),
\end{equation}
where $X_{,\mu} $ denotes $\partial_\mu X$.

Weyl's gauge (scale) transformation means that the line element $ds^2 =
g_{\mu\nu}dx^\mu dx^\nu $ is transformed under the rule,
\begin{equation}
  ds^2 \to ds'^2 = \Omega^2 (x) ds^2 .
\end{equation}
If we choose to keep the parametrization of coordinates $x^\mu$ invariant,
Weyl's gauge transformation can be equivalently represented by the conformal
transformation of the metric,
\begin{equation}
  g_{\mu\nu}(x) \to g'_{\mu\nu}(x) = \Omega^2 (x) g_{\mu\nu}(x).
\end{equation}
It is clear that Weyl's gauge transformation is not included in Einstein's 
general transformation of coordinates, under which $ds^2 $ is an invariant. 
Also, the light-speed still keeps the unity in Weyl theory since Weyl's gauge 
transformation has nothing to do with null geodesics $ds^2=0$. 
Generally, a quantity $X$ would be said to have conformal weight $n$ if it
transforms as
\begin{equation}
  X \to X' = \Omega^n (x) X.
\end{equation}

Suppose $\Omega (x)$ is a constant of space-time, then the Christoffel symbol
would be invariant under this transformation. However, if 
$\Omega (x)$ is space-time dependent, to keep 
$\Gamma_{\mu\rho}^\nu $ conformal invariant would require an additional
{\it gauge} field $S_\mu $ transformed as
\begin{equation}
  S_\mu \to S'_\mu = S_\mu - \Omega^{-1}(x)\partial_\mu\Omega (x), 
  \label{gaugtrans}
\end{equation}
and the partial derivative in the Christoffel symbol replaced by the 
conformal covariant derivative
\begin{equation}
  {\cal D}_\mu g_{\nu\rho } \equiv (\partial_\mu + 2S_\mu ) g_{\nu\rho }.
\end{equation}
Hence the conformal invariant affine connection is given by
\begin{eqnarray}
  \tilde{\Gamma}_{\mu\rho}^\nu &\equiv& {1\over 2}g^{\nu\sigma}\left( 
    {\cal D}_\rho g_{\sigma\mu} + {\cal D}_\mu g_{\sigma\rho} - 
    {\cal D}_\sigma g_{\mu\rho} \right) \nonumber\\
  &=& \Gamma_{\mu\rho}^\nu + \delta^\nu_\mu S_\rho
    + \delta^\nu_\rho S_\mu - g_{\mu\rho} S^\nu ,
\end{eqnarray}
and the corresponding scalar curvature becomes
\begin{equation}
  \tilde{R} = R + 6 D_\mu S^\mu + 6 S_\mu S^\mu .
\end{equation}
Note that ${\cal D}_\mu g_{\nu\rho }$ and $\tilde{R}$ have weights $2$ 
and $-2$ respectively. Furthermore, the conformal covariant derivative
of a vector field $V^\nu$ with weight $n$ is written as  
\begin{equation}
  {\cal D}_\mu V^\nu \equiv (\partial_\mu + n S_\mu )V^\nu
    -\tilde{\Gamma}_{\mu\rho}^\nu V^\rho .
\end{equation}

It immediately follows that all parameters in any gravity theory with scale 
invariance are dimensionless. In order to have Einstein gravity as the 
effective theory at low energies, the dimensional "constants" in the 
Einstein gravity such as 
the gravitational constant and the cosmological constant must correspond to 
quantities composed of some fields with dimension, whose values may also be
changing with space-time. To implement this, we introduce an additional scalar 
field $\hat{\phi}$ with weight $-1$. A self-interaction term of $\hat{\phi}$, 
$V(\hat{\phi})$, can also be added as long as it is being scale invariant:
\begin{equation}
 {\delta\over\delta\Omega}\sqrt{-g'}V(\hat{\phi}') 
= \sqrt{-g'}\Omega^{-1}\left[ 4 V(\hat{\phi}') - 
     \hat{\phi}'{\delta V(\hat{\phi}') \over \delta\hat{\phi}' }\right] = 0,
\label{reqscalinv}
\end{equation}
which simply means that the stress-energy tensor of $\hat{\phi}$ must be
traceless. The only non-trivial choice of $V$ consistent with the above 
equation is 
\begin{equation}
  V(\hat{\phi}) = {\lambda\over 4!}\hat{\phi}^4,
\end{equation}
up to a coupling constant $\lambda$.

Thus, we can construct a simple scale invariant action for gravity as 
\begin{equation}
  S_{(\xi>0)}=\int d^4 x\sqrt{-g}\left[ -{1\over 2}\xi\hat{\phi}^2 \tilde{R}
    -{1\over 2}{\cal D}_\mu \hat{\phi} {\cal D}^\mu \hat{\phi}
    - V(\hat{\phi}) - {1\over 4} H_{\mu\nu} H^{\mu\nu}\right] ,
\label{weylL}
\end{equation}
in which the signature is $(-,+,+,+)$, and
\begin{eqnarray}
  \tilde{R} &=& R + 6f(D_\mu + f S_\mu )S^\mu ,\\
  {\cal D}_\mu \hat{\phi} &=& (\partial_\mu - fS_\mu )\hat{\phi},\\
  H_{\mu\nu} &=& \partial_\mu S_\nu - \partial_\nu S_\mu ,
\end{eqnarray}
where $f$ is a coupling constant, and $\xi>0$ to allow for a positive
gravitational constant.
The action is invariant under the local scale transformations,
\begin{eqnarray}
  g_{\mu\nu}(x) &\to & g'_{\mu\nu}(x)=\Omega^2(x) g_{\mu\nu}(x),\label{scale1}\\
  \hat{\phi}(x) &\to & \hat{\phi}'(x)=\Omega^{-1}(x) \hat{\phi}(x),
    \label{scale2}\\
  S_\mu (x) &\to & S'_\mu (x) = S_\mu (x) - f^{-1}\Omega^{-1}(x)
    \partial_\mu\Omega (x).\label{scale3}
\end{eqnarray}
In fact, this action is an usual form of the so-called classical scale
invariant gravity theory which has been extensively studied in the
literatures\cite{dirac,utiyama}.

\section{Generalized Scale Invariant Gravity}

Before we study the mass spectrum of the Lagrangian in the action 
$S_{(\xi>0)}$~$(\ref{weylL})$, let us rewrite it in a more convenient 
form,
\begin{equation}
  L_{(\xi>0)}=-\sqrt{-g}\left[ {1\over 2}\xi\hat{\phi}^2 R
   +{1\over 2} \partial_\mu \hat{\phi} \partial^\mu \hat{\phi}
   +{1\over 2} (1+6\xi) f \hat{\phi}^2 (f S_\mu S^\mu + D_\mu S^\mu) 
   + {1\over 4} H_{\mu\nu} H^{\mu\nu} + V(\hat{\phi}) \right].
\label{lag1}
\end{equation}
Suppose the true vacuum state be the 
Minkowski flat space-time obeying Einstein gravity, then we perturb the classical fields about this vacuum,
\begin{eqnarray}
  g_{\mu\nu} &=& \eta_{\mu\nu} + h_{\mu\nu},\\
  {1\over 2} \xi \hat{\phi}^2 &=& \frac{1}{16\pi G} e^{\alpha\sigma}, \\
  S_\mu &=& s_\mu,
\end{eqnarray}
where $\alpha$ is an arbitrary parameter, and $G$ is the Newton's constant.
Also, all $h_{\mu\nu}$, $\sigma$, 
and $s_\mu$ are small perturbations. After redefining a new metric 
perturbation,
\begin{equation}
  \rho_{\mu\nu} = h_{\mu\nu} +\alpha\sigma\eta_{\mu\nu},
\end{equation}
the kinetic part in $L_{(\xi>0)}$ can be expanded up to quadratic terms
into
\begin{eqnarray}
{\cal L}_{K.E.} &=& -\frac{1}{16\pi G} \left\{
  {1\over 4} \partial_\mu\rho_{\alpha\beta}\partial^\mu\rho^{\alpha\beta}
 -{1\over 4} \partial_\mu\rho\partial^\mu\rho
 +{1\over 2} \partial_\alpha\rho^{\alpha\beta}\partial_\beta\rho
 -{1\over 2} \partial_\alpha\rho^{\alpha\mu}\partial^\beta\rho_{\beta\mu}
 \right.\nonumber\\& &\left. +\frac{1+6\xi}{\xi} \left[ 
    \frac{\alpha^2}{4}\partial_\mu \sigma \partial^\mu \sigma
   +f^2 s_\mu s^\mu + f \partial_\mu s^\mu \right] \right\}
 -{1\over 4} (\partial_\mu s_\nu - \partial_\nu s_\mu)
             (\partial^\mu s^\nu - \partial^\nu s^\mu),
\label{lagk}
\end{eqnarray}
where $\rho=\eta^{\mu\nu}\rho_{\mu\nu}$.
To maintain the stability of the theory, one has to assure the positivity
of the kinetic energy as well as the mass of each perturbation. This requires 
that $(1+6\xi)/\xi$ must be greater than zero, i.e., either $\xi>0$ or
$\xi \le -1/6$. We thus see that the stability condition does not rule out
a negative value for $\xi$. However, that $\xi \le -1/6$ would imply an 
unwanted negative gravitational constant. Therefore, in order to include a
negative $\xi$, we propose the following generalized scale invariant 
action for gravity,
\begin{equation}
  S_\xi=\int d^4 x\sqrt{-g}\left[ -{1\over 2}|\xi|\hat{\phi}^2 \tilde{R}
    -\frac{|\xi|}{\xi}\left(
    {1\over 2}{\cal D}_\mu \hat{\phi} {\cal D}^\mu \hat{\phi}
    + V(\hat{\phi}) \right)
    - {1\over 4} H_{\mu\nu} H^{\mu\nu}\right],
\label{act}
\end{equation}
where $\xi$ is any real number excluding the interval $(-1/6,0]$.  
Obviously, this action reduces to $S_{(\xi>0)}$ when $\xi>0$. 
Note that the kinetic energy of $\hat{\phi}$ in the action $(\ref{act})$ 
would be negative when $\xi \le -1/6$. This may bring out an energy
crisis at the classical level, even though we have just shown that the 
positivity of energy is guaranteed in the spectrum of the theory. 
However, we will show in Section V that the total energy of the 
classical theory is bounded below and well-defined.

When $\xi = -1/6$, the gauge field $S_\mu$ in the action $(\ref{act})$ 
becomes massless and decouples from the scalar field $\hat{\phi}$, namely,
\begin{eqnarray}
  S&=& S_{grav}+S_{gauge};\\
  S_{grav}&=& \int d^4 x \sqrt{-g}\left[ -{1\over 12}\hat{\phi}^2 R + 
     {1\over 2}\partial_\mu\hat{\phi}\partial^\mu\hat{\phi}
     +V(\hat{\phi})\right],\label{action}\\
  S_{gauge}&=&\int d^4 x \sqrt{-g}\left[ -{1\over 4}H_{\mu\nu}H^{\mu\nu}\right].
\end{eqnarray}
In this case the resulting action $S_{grav}$ describes a theory of
a scalar field $\hat{\phi}$ conformally coupled to gravity,
which is invariant under the local scale transformations 
$(\ref{scale1})$ and $(\ref{scale2})$. Note that $S_{gauge}$
is by itself invariant under the transformation $(\ref{scale3})$ with the transformation function $\Omega(x)$ independent of the ones for $S_{grav}$. 
Hence the dynamical degrees of freedom (DOF) of the $\xi =-1/6$ theory is less than those of the cases with $\xi\not= -1/6$ by one. For $\xi\not= -1/6$, the
dynamical DOF is nine: ten DOF for graviton, one for scalar field and four for
gauge field altogether amount to fifteen DOF,  which is then deduced by one
scale symmetry, four coordinate reparametrization symmetry 
and a constraint Gauss' law. And for $\xi =-1/6$, the dynamical DOF is eight.
Also, from Eq.~$(\ref{lagk})$, we see that 
$\hat{\phi}$ is in fact an auxiliary field in the spectrum. This is a 
manifestation of the scale symmetry.
To fix the degree of freedom corresponding to the scale symmetry,
one can put a constraint, the {\it Einstein gauge},
\begin{equation}  
  \hat{\phi} = v, \label{Eingau0}
\end{equation}
with $v$ being a non-zero constant, or equivalently,
\begin{equation}
  \Omega^2(x) = v^{-2}\hat{\phi}^2, \label{Eingau}
\end{equation}
so that the action $S_{grav}$ becomes
\begin{equation}
  S_{EG}=\int d^4 x \sqrt{-g}\left[ -{1\over 16\pi G}R +2\Lambda \right],
\label{EG}
\end{equation}
which is identical to the Einstein gravity with the gravitational constant
$G=3/(4\pi v^2)$ and the cosmological constant $\Lambda =\lambda v^4/48$.
We thus see that the so-called Weyl gravity $(\ref{action})$ is no more 
than a generalized form of the Einstein gravity $S_{EG}$. 
In general, it represents a set of 
pure gravity theories with arbitrary space-time varying gravitational and
cosmological "constants" interconnected by the local scale transformation. 

Below we will restrict our calculations to the massless case $(\ref{action})$
for simplicity. Technically, this suffices to show how we would deal with the
local scale symmetry in doing quantum corrections in the next section.
The method can be equally well applied to the massive cases. It is rather
interesting to see how the massless case can be related to the massive cases
through a running coupling constant $\xi$.

\section{One-Loop Effective Potential}

To study the quantum aspect of the theory $(\ref{action})$, 
it is convenient to start with the 
loop-corrected effective action $\cite{schwinger,CW}$.
Since we are mainly interested in using the quantum corrected result to 
study the ground state of the theory and also their implications to classical 
cosmology, we are going to ignore the singular behavior of the system. Thus
it is appropriate to use the perturbation theory, in which the gravity is 
depicted as a theory of massless spin-two gauge particles, namely, 
the gravitons. However, there exist some ambiguities when one applies the
conventional Coleman-Weinberg formalism to gauge theories. 

The first ambiguity is that the conventional effective action is not invariant
under field reparametrizations or, in particular, gauge transformation.
Although technically the gauge invariance of the effective action can be 
achieved by splitting each field into quantum- and background-part, and making 
an arrangement so that the gauge of the quantum-field-part is fixed while the 
effective action composed by the background-field-part is
gauge-invariant$\cite{abbott}$, it is not garanteed that the same effective
action can be obtained if one initially chooses another gauge condition or 
another field parametrization for quantum fields. 

Even if a gauge invariant effective potential is worked out, usually it still 
remains a dependence on the coupling factor of the gauge fixing term. 
This factor gives rise to another ambiguity, especially in direct applications
of the effective potential in practical situations. Fortunately, the whole
problem was resolved by Vilkovisky$\cite{vilkov}$ and DeWitt$\cite{dewitt}$.
Recalling Einstein's idea in formulating the theory of general relativity: 
the coordinate reparametrization dependence can be eliminated by simply 
choosing all variables and derivatives to be covariant, Vilkovisky and DeWitt 
pointed out that the space of the field configurations may not be
trivially flat. Thus, in order to obtain an effective action independent to the
field reparametrization, one should properly construct covariant variables and 
derivatives in the configuration space. Furthermore, it is possible to define a
gauge invariant metric in the configuration space, so that a gauge independent effective potential can be uniquely derived without any ambiguity.

As the effective potential is taken from the zeroth order of the momentum 
expansion of the effective action, the effective potential is off-shell 
except at the vacuum, where the field configuration satisfies the equation 
of motion. So, it is not unexpected that 
the shape of the effective potential is gauge-dependent. However, 
the order of the phase transition
during the symmetry breaking process depends on the shape of the effective
potential, and a gauge-independent effective potential is needed in
computing thermal quantities sensitive to the order of phase transition.
In addition, a gauge-independent effective potential is useful
in some cases in which the existence of the imaginary part of the effective
potential depends on gauge choices\cite{naive}.

As such, we will adopt the Vilkovisky and DeWitt's (VD) method to calculate 
the loop corrections. Instead of displaying the full VD calculation of the 
effective potential, we begin with the conventional Coleman-Weinberg formalism
and modify it to the VD calculations when necessary. A brief account of the VD
method and a detailed calculation of the effective potential are given in
Appendix A.

Expanding the gravitational field and the scalar field about the 
ground-state background, one has
\begin{eqnarray}
  g_{\mu\nu} &=& \eta_{\mu\nu} + \phi^{-1} h_{\mu\nu},\label{BF1}\\
  \hat{\phi} &=& \phi + \sigma,\label{BF2}
\end{eqnarray}
where $\eta_{\mu\nu}$ is a flat metric, $\phi$ is a constant field, 
and the quantum field $h_{\mu\nu}$ is graviton.
In order to make the generators of the transformation of the scalar field 
dimensionless, one may define
\begin{equation}
  \tilde{\phi} \equiv \phi^{-1} \hat{\phi} = 1+\phi^{-1} \sigma.
\end{equation}
To evaluate the one-loop effective potential, it suffices to expand
the action $(\ref{action})$ up to the second order in the quantum fields, 
namely,
\begin{eqnarray}
  -\sqrt{g}{\cal L} \approx {\cal L}_2 &\equiv& 
    V(\phi)+{V(\phi)\over 2\phi}h+V'(\phi)\sigma\nonumber\\
&-& {1\over 24}\left[ h^{\mu\nu}{}_{,\mu}h_{,\nu}-{1\over 2}h_{,\mu}h^{,^\mu}
    -h^{\mu\nu}{}_{,\nu}h_{\mu\rho,}{}^{\rho} 
    +{1\over 2}h^{\mu\nu,\rho}h_{\mu\nu,\rho} 
    -4\sigma ( h_{,\mu}{}^{,\mu}-h^{\mu\nu}{}_{,\mu\nu} )\right] \nonumber\\
&+& {1\over 2}\sigma_{,\mu}\sigma^{,\mu} + {1\over 8}{V\over \phi^2}h^2
    +{1\over 2}{V'\over \phi}h\sigma +{1\over 2}V''\sigma^2
    -{1\over 4}{V\over \phi^2}h^{\mu\nu}h_{\mu\nu},\label{L2}
\end{eqnarray}
where $h\equiv h_\mu{}^\mu$ and $V'\equiv {\delta V/ \delta\phi}$.
One may write the quadratic part of the Lagrangian in the following form,
\begin{equation}
  {\cal L}_q = {1\over 2}\psi_a P^{ab} \psi_b,\label{simpleL}
\end{equation}
where $a,b = 0,\cdots ,10 $ and $\psi_a$ represent the quantum fields
$\sigma$ and ten independent components of $h_{\mu\nu}$ respectively. 
 
Suppose a transformation
\begin{equation}
  \rho_{\mu\nu} = h_{\mu\nu} +2\sigma\eta_{\mu\nu}\label{infscale}
\end{equation}
is performed, the quadratic Lagrangian becomes
\begin{eqnarray}
  {\cal L}_2 &=&-{1\over 24}\left( {1\over 2}\rho\partial^2\rho
    +\rho_{\mu\nu}\partial^\nu\partial_\beta\rho^{\mu\beta} 
    -\rho\partial_\mu\partial_\nu\rho^{\mu\nu}  
    -{1\over 2}\rho_{\mu\nu}\partial^2\rho^{\mu\nu}\right) + {V\over 4\phi^2}
    \left( {1\over 2}\rho^2-\rho_{\mu\nu}\rho^{\mu\nu} \right)\nonumber\\
  & & +\sigma A \rho + \sigma B \sigma,\label{Lrho}\\
  A &\equiv& {V'\over 2\phi}-{V\over\phi^2},\nonumber\\ 
  B &\equiv& {V''\over 2}-{4V'\over\phi}+{4V\over \phi^2},\nonumber
\end{eqnarray}
which is equivalent to the Lagrangian $(\ref{lagk})$ with $\xi =-1/6$, 
where we have performed a different expansion scheme in order to make the gravitational constant $G$ explicit.
Here the kinetic term of $\sigma$ field vanishes, that is, $\sigma$ is an 
auxiliary field. This gives a hint that there are some symmetries in this Lagrangian. Indeed, they originate from the scale transformations
$(\ref{scale1})$ and $(\ref{scale2})$. When 
the Einstein gauge $(\ref{Eingau})$ is chosen, the gravitational field becomes
\begin{equation}
  g'_{\mu\nu} = v^{-2} \hat{\phi}^2 g_{\mu\nu}.
\end{equation}
Substituting the background field expansions $(\ref{BF1})$ and $(\ref{BF2})$
into the above transformation, one would obtain the infinitesimal version 
of the scale transformation,
\begin{equation}
  \rho_{\mu\nu}\equiv h'_{\mu\nu}=h_{\mu\nu}+2\sigma\eta_{\mu\nu}+O(\psi^2),
\end{equation}
if $v = \phi$. This is exactly the transformation $(\ref{infscale})$.

One can go further by letting
\begin{equation}
  \sigma' =\sigma +{A\over 2B}\rho 
    =\left(1+{4A\over B}\right)\sigma+{A\over 2B}h,
\end{equation}
so that the Jacobian with respect to the field reparametrization 
$(h_{\mu\nu},\sigma)\to (\rho_{\mu\nu},\sigma ')$ equals unity.
Then the last two terms in Eq.~$(\ref{Lrho})$ turn into
\begin{equation}
  \sigma A\rho +\sigma B\sigma = \rho\left({A^2\over 4B}\right)\rho +
                                 \sigma' B\sigma'\label{egwg},
\end{equation}
where $\sigma' $ is now decoupled. 
The field equation $\sigma' =0$ corresponds to the
Einstein gauge $(\ref{Eingau0})$. We thus conclude that the only difference
between the Einstein gravity and the Weyl gravity $(\ref{action})$ is that 
the latter has the gravitational wave with a mass term proportional
to $\rho^2 $ as shown in the right-hand side of Eq.~$(\ref{egwg})$. 
Nevertheless, this difference can be lifted by choosing a traceless 
gauge $\rho=0$.

In fact, after the Lagrangian $(\ref{simpleL})$ being diagonized,
there are other vanishing kinetic terms, which correspond to
four degrees of freedom of the reparametrization symmetry. 
This property is not surprising in systems with internal symmetries.
Consider a symmetry transformation operator $U$ operating on a quantum
field by $\psi_a \to \psi'_a = U_a{}^b \psi_b$. Assume the 
quadratic Lagrangian is invariant under such a symmetry transformation, 
then $P = U^T PU$ where $U^T$ denotes the transverse of $U$.
This would imply that det$P =0$ if det$U\not= 1$.
In our model $(\ref{L2})$, the determinants of the operators corresponding 
to the scale transformation and reparametrization are indeed not equal to unity.
As such, the corresponding operator $P^{ab}$ $(V=0)$ is not invertible, and 
hence its propagator cannot be defined. Similar situations occur in the 
theory of electromagnetism, where one can interpret $P$ as 
a projection operator$\cite{kaku}$.

It is therefore preferable to set the gauge condition in a form of a first
derivative with respect to the corresponding quantum field in this case. 
For example, in electromagnetism, 
one may choose the Lorentz gauge $\partial_\mu A^\mu =0$,
whose corresponding gauge fixing term reads
\begin{equation}
  {1\over 2\alpha}(\partial_\mu A^\mu )^2
\end{equation}
with an arbitrary factor $\alpha$. Once this term is added
into the Lagrangian of electromagnetism, the propagator of photon is then 
well-defined up to the factor $\alpha$.

In the present case, the reparametrization symmetry can be fixed by choosing
\begin{equation}
  h_{\mu\nu,}{}^{\nu}-{1\over 2}h_{,\mu} =0,
\end{equation}
while the gauge condition for the scale symmetry $(\ref{Eingau0})$ is 
equivalent to
\begin{equation}
  \partial_\mu \sigma =0.
\end{equation}
Hence, the gauge fixing term then reads
\begin{equation}
  {\cal L}_{gf} = {1\over 2\alpha} (h_{\mu\rho,}{}^\mu - {1\over 2}h_{,\rho})
    (h^{\nu\rho}{}_{,\nu}-{1\over 2} h_,{}^\rho) 
    + {1\over 2\beta} \sigma_{,\mu}\sigma_,{}^\mu,
\end{equation}
where $\alpha$ and $\beta$ are arbitrary factors.

In the one-loop level, only quadratic terms are needed in path-integral 
calculation. Since the corresponding ghost Lagrangian,
\begin{equation}
  {\cal L}_{ghost} = \bar{\eta}_\mu(-\partial^2)\eta^\mu,
\end{equation}
is totally decoupled from the system in this level, the ghost field can 
be neglected here. The quadratic part of the Lagrangian considered in 
this approximation is
\begin{eqnarray}
  {\cal L}_q &=& 
    {1\over 4}h_{\mu\nu}\alpha_1 h^{\mu\nu} -{1\over 4}h\alpha_2 h
   +{1\over 2}h_{\mu\nu}\alpha_3{\partial^\nu\partial_\rho\over\partial^2}
     h^{\mu\rho}
   -{1\over 2}h\alpha_4{\partial_\mu\partial_\nu\over \partial^2}h^{\mu\nu}
   -{1\over 2}h_{\mu\nu}\alpha_5
    {\partial^\mu\partial^\nu\partial_\rho\partial_\sigma\over\partial^4}
    h^{\rho\sigma}\nonumber\\
   &+& {1\over 2}\sigma\beta_1\sigma +{1\over 2}\sigma\beta_2 h
   +{1\over 2}\sigma\beta_3{\partial_\mu\partial_\nu\over \partial^2}
    h^{\mu\nu},\label{Lq}
\end{eqnarray}
where
\begin{eqnarray}
  \alpha_1 &=& -{k^2\over 12}-{\lambda\phi^2\over 24},\nonumber\\
  \alpha_2 &=& -\left( {1\over 12} +{1\over 2\alpha}\right) k^2 -
               {\lambda\phi^2\over 48},\nonumber\\
  \alpha_3 &=& \alpha_4 = \left( {1\over 12}+{1\over \alpha} \right)k^2,
               \nonumber\\
  \alpha_5 &=& 0,\nonumber\\
  \beta_1 &=& \left( 1+{1\over\beta}\right) k^2 +{\lambda\phi^2\over 2},
              \nonumber\\  
  \beta_2 &=& {k^2\over 3}+{\lambda\phi^2\over 6},\nonumber\\
  \beta_3 &=& -{k^2\over 3},
\end{eqnarray}
with a replacement $-\partial^2\to k^2$. Here $\partial^4 $ is a shorthand of
$(\partial_\mu\partial^\mu)^2$. Let us write ${\cal L}_q$ in the form of 
Eq.~$(\ref{simpleL})$, the eigenvalues of $P^{ab}$ are found to be
\begin{eqnarray}
  \lambda_1&=&\lambda_2=\lambda_3=-{k^2\over 12}-{\lambda\phi^2\over 24},
               \nonumber\\
  \lambda_4&=&\lambda_5=\lambda_6={k^2\over\alpha}-{\lambda\phi^2\over 24},
               \nonumber\\
  \lambda_7&=&\lambda_8=-{k^2\over 24}-{\lambda\phi^2\over 48},\nonumber\\
  \lambda_0\lambda_9\lambda_{10} &=& {k^6\over 48\alpha\beta} 
    +\lambda\phi^2\left( {1\over 96\alpha\beta}-{1\over 48\alpha}
     -{1\over 1152\beta}\right) k^4\nonumber\\ & &
    +\left( \lambda\phi^2\right)^2 \left( {1\over 3456}-{1\over 64\alpha}
     -{1\over 2304\beta}\right) k^2
    +{5\over 13824}\left(\lambda\phi^2\right)^3.\label{conven}
\end{eqnarray}
Indeed, if $1/\alpha$ and $1/\beta$ are set to be zero such that 
${\cal L}_{gf}$ vanishes, there would be five elements of the eigenvector
which have no kinetic terms. They are $\lambda_4$,$\lambda_5$, $\lambda_6$ 
and two of the three eigenvalues $\lambda_0$, $\lambda_9$ and $\lambda_{10}$.

In terms of $\lambda$'s, the unrenormalized one-loop effective potential
can be written as
\begin{equation}
  V_1 = V - {i\over 2} \sum_{a=0}^{10} {\rm Tr}\ln\lambda_a.\label{1loop}
\end{equation}
Obviously, the conventional effective potential obtained by substituting
the eigenvalues $(\ref{conven})$ into Eq.~$(\ref{1loop})$ depends on
arbitrary factors $\alpha$ and $\beta$. To eliminate this ambiguity, one
should introduce the Vilkovisky-DeWitt effective potential.

>From Appendix A, the Vilkovisky-DeWitt method changes the eigenvalues into
\begin{eqnarray}
  \lambda_1=\lambda_2 &=& \lambda_3=-{k^2\over 12}-{\lambda\phi^2\over 24},
            \nonumber\\
  \lambda_4 = \lambda_5 &=& \lambda_6 = {k^2\over \alpha},\nonumber\\
  \lambda_7 = \lambda_8 &=& -{k^2\over 24}-{\lambda\phi^2\over 48},\nonumber\\
  \lambda_0\lambda_9\lambda_{10} &=&{k^4\over 48\alpha\beta}
    \left( k^2 + {505\over 1058}\lambda\phi^2 \right),\label{L910}
\end{eqnarray}
by combining the original quadratic Lagrangian $(\ref{Lq})$ with the 
correction $(\ref{L'})$. The one-loop VD effective potential then reads
\begin{equation}
  V_1^{VD} = {\lambda\phi^4\over 4!} +
    {5i\over 2}{\rm Tr}\ln\left( k^2+{1\over 2}\lambda\phi^2\right)+
    {i\over 2}{\rm Tr}\ln\left( k^2+{505\over 1058}\lambda\phi^2\right)
    + {\rm constant}.
  \label{nonrenormVD}
\end{equation}
This result is similar to those obtained from the simple massless $\phi^4$
theory.
Following the standard renormalization process$\cite{kaku}$, the renormalized
VD effective potential can be calculated as
\begin{eqnarray}
  V_1^{VD} &=& {\lambda \phi^4\over 4!} - {(\lambda \phi^2)^2\over 64\pi^2}
    K \left( \ln{\phi^2\over M}-{25\over 6}\right) + \Lambda
    +O(\lambda^3),\label{V1VD}\\
  K&=& {5\over 4}+\left( {505\over 1058}\right)^2,
\end{eqnarray}
where $M$ is a scaling factor, and $\Lambda$ is a renormalized constant 
which will be determined in the following section. If we simply choose the 
Landau-DeWitt gauge, $\alpha=\beta=0$, in the conventional effective potential
obtained from the eigenvalues $(\ref{conven})$, we would
obtain a slightly different value of $K=3/2$.  

Note that the renormalized effective potential $(\ref{V1VD})$ would have no
imaginary part if $\lambda$ is positive. In contrary, in the massive case 
$(\ref{weylL})$\cite{smolin}, the imaginary part of the conventional effective
potential under the Laudau-DeWitt gauge is nonzero as long as $\lambda >0$,
while it vanishes when $\lambda <0$. Moreover, some other gauge choices 
may make the imaginary part always exist. 
Since $\lambda$ has to be positive
in order to allow for a stable vacuum, our results 
suggest that the coupling constant $\lambda |\xi | /\xi$ of the self- 
interaction in the action $(\ref{act})$ should be always negative.

Here we have some remarks and comments on the VD method. First, it is obvious 
that there is no special values of $\alpha $ and $\beta $ be chosen such 
that $V_1 =V_1^{VD}$ in this scale invariant gravity. 
Second, the effect of VD method on the eigenvalues $\lambda_4$, $\lambda_5$ 
and $\lambda_6$ is to remove the vertex term $-\lambda\phi^2 /24$
(see Eqs.~$(\ref{conven})$ and $(\ref{L910})$), so that the 
loops of $\psi_4$, $\psi_5$ and $\psi_6$ are decoupled from the system. 
This decoupling can also be achieved by naively  
choosing the Landau-DeWitt gauge. However, 
it is not the case in calculating $\lambda_0\lambda_9\lambda_{10}$. 
The complicated mixing between these quantum fields makes the VD residual 
non-zero vertex term different from that obtained from the 
Landau-DeWitt gauge. In general, the equality between the naive and the VD
effective potentials occurs only in some special cases by accident.

Recall that when a constrained or gauge system is quantized in the path 
integral formalism, the constraints or gauge conditions $\delta (F[\phi ])$ 
are loosed into a Gaussian distribution $e^{-F^2/2\alpha}$, where $\alpha $ 
can be defined as the width of this distribution. In other words, 
all of the off-shell field configurations, which are weighted by a Gaussian, 
are taken into account, and the constraint $F[\phi ]=0$ is true only when the
system is on-shell. Therefore, to choose the Landau-DeWitt gauge, $\alpha\to 0$,
is equivalent to narrowing the Gaussian distribution to a delta-function-like 
distribution. However, if the configuration space is curved, the direction 
that $\alpha\to 0$ may not be orthogonal to the on-shell surface $F[\phi ]=0$
everywhere because $\alpha $ is not a covariant quantity in ${\cal M}$.
Hence taking the Landau-DeWitt gauge naively without considering the curvature 
effect may give the wrong result. 

For example, if we choose the Landau-DeWitt gauge in the conventional effective
potential obtained from Eq.~$(\ref{conven})$, the resulting effective potential
is identical to the conventional effective potential obtained from Einstein 
gravity $(\ref{EG})$ with $\alpha\to 0$. It should be emphasized that the 
Einstein gravity is the consequence of choosing the gauge, 
$\sigma =0$, in the scale invariant gravity $(\ref{Lrho})$ {\it before} 
quantization. However, neither of them is equal 
to the VD effective potential $(\ref{nonrenormVD})$. Although the Einstein
gravity and the scale invariant gravity are classically equivalent, the
off-shell structure as well as the quantum theory of them are quite different.

The third remark is that the field $\sigma$, which has no kinetic term, can be 
identified as an auxiliary field in the classical theory. One may substitute 
the equation of motion with respect to $\sigma$ in the Lagrangian to eliminate
the auxiliary field and get a new pure graviton theory. If we 
include only the graviton gauge-fixing $\alpha$-term to 
compute the one-loop VD effective potential for this new theory, 
the results would depend on $\alpha$.
This is because we have ignored the scale symmetry hidden in the new theory.  
Therefore, if one finds that the obtained VD effective potential 
still depends on some arbitrary factor which corresponds to the known
symmetry of a system, then the system would have to carry some extra hidden
symmetry. In this case, one may apply the method developed by 
Dirac$\cite{Dirac2}$ to find all the constraints and then run the 
quantization process again.

\section{Vacuum Expectation Value}

The vacuum expectation value of $\phi$ is located at the minimum of 
$V_1^{VD}$, namely,
\begin{equation}
  \left. {\delta V_1^{VD}\over \delta\phi }\right|_{\left<\phi\right>} =0.
\end{equation}
From Eq.~$(\ref{V1VD})$, this implies that 
\begin{equation}  
  \left<\phi\right> = \sqrt{M}\exp \left[-{4\pi^2\over 3\lambda K}
    +{11\over 6} \right].
\end{equation}
For every non-zero coupling constant $\lambda$, the vacuum expectation
value $\left< \phi\right> $ does not vanish. This is one of the characteristics 
of the symmetry breaking. Note that not only the scale symmetry of the ground 
state but also that of the Lagrangian are broken. In fact, the scale symmetry 
was broken manifestly in the process of doing renormalization: on the
onset a mass counterterm has been introduced in the total Lagrangian. Such a 
symmetry-breaking would result in a trace-anomaly. However, the scale symmetry 
survives at the false vacuum $\phi =0$.

Replacing the scaling factor $M$ by $\left<\phi\right> $, the effective 
potential can be written as
\begin{equation}
  V_1^{VD} = {K\lambda^2\phi^4\over 64\pi^2}\left( 
    2\ln {\phi\over\left<\phi\right> } -{1\over 2}\right) + \Lambda.
\end{equation}
We would choose 
\begin{equation}
  \Lambda={K \lambda^2 \left<\phi\right> ^4 \over 128\pi^2}
\end{equation}
so that $\left. V_1^{VD}\right| _{\left<\phi\right> } =0$.

Let us consider a cosmological model 
including the classical matter Lagrangian 
${\cal L}_M$ and the effection action of the scale invariant gravity:
\begin{equation}
  S= \int d^4 x \sqrt{-g}\left[ -{1\over 12} \phi^2 R + {1\over 2}
   \partial_\mu\phi\partial^\mu\phi + V_1^{VD}(\phi)+{\cal L}_M \right] .
\label{actM}
\end{equation}
We obtain the field equations,
\begin{eqnarray}
  -{\phi^2\over 6}\left( R^{\mu\nu}-{1\over 2}R g^{\mu\nu}\right) &=&
   -\phi^{;\mu}\phi^{;\nu}+{1\over 2}g^{\mu\nu}\phi_{;\rho}\phi^{;\rho}
   -{1\over 6}g^{\mu\nu}(\phi^2)_{;\rho}{}^{;\rho}
   +{1\over 6}(\phi^2)^{;\mu;\nu}+g^{\mu\nu}V_1^{VD}(\phi)+T_M^{\mu\nu},
   \label{eom1}\\
  0 &=& \phi_{;\mu}{}^{;\mu} + {1\over 6}R\phi - V_1^{VD}{}'(\phi),
   \label{eom2}
\end{eqnarray}
by varying the action with respect to $g_{\mu\nu}$ and $\phi$.
Here the stress-energy tensor of classical matter is defined by
\begin{equation}
  T_M^{\mu\nu} \equiv 2{\delta\over\delta g_{\mu\nu}}
                        \int d^4 x\sqrt{-g} {\cal L}_M. 
\end{equation}
Comparing the covariant derivative of Eq.~$(\ref{eom1})$ 
with Eq.~$(\ref{eom2})$, one has the energy-momentum conservation law 
of the classical matter,
\begin{equation}
  T_M^{\mu\nu}{}_{;\nu} = 0.\label{conserv}
\end{equation}
Further, taking trace on Eq.~$(\ref{eom1})$ then comparing with 
Eq.~$(\ref{eom2})$, a simple but strong constraint,
\begin{equation}
 4V_1^{VD}(\phi)-\phi V_1^{VD}{}'(\phi)  = -T_M \equiv -g_{\mu\nu}T_M^{\mu\nu}.
 \label{key}
\end{equation}
can be obtained. Note that when $T_M$ vanishes, the above equation is exactly
the constraint $(\ref{reqscalinv})$ required by the scale symmetry.

In standard cosmology$\cite{Kolb}$, the stress-energy tensor of  
classical matter can be approximated by a perfect fluid with  
energy density $\rho$ and pressure $p$:
\begin{equation}
  T^{\mu\nu}= p g^{\mu\nu} + (\rho +p)U^\mu U^\nu ,
\end{equation}
where $U^\mu $ is a four-velocity with $U^\mu U_\mu = -1$.
Assume a homogeneous and isotropic universe, 
then $\rho$, $p$ and $\phi$ are functions of time only, 
and the geometry of the universe is given by
the Friedmann-Robertson-Walker metric,
\begin{equation}                        
  ds^2 = -dt^2 + a^2(t)\left[ {dr^2\over 1-kr^2}+r^2 d\Omega^2\right],
  \label{frw}
\end{equation}
where $a(t)$ is the cosmic scale factor,
and $k=1,0$ or $-1$ corresponds to a closed, flat or open universe 
respectively. Inserting the metric into Eq.~$(\ref{eom1})$, one obtains
\begin{eqnarray}
  {\phi^2\over 2a^2}\left(\dot{a}^2+k\right) &=& -{1\over 2}\dot{\phi}^2
    -{1\over 2}{\dot{a}\over a}\dot{\left(\phi^2\right)}
    +\rho-V_1^{VD},\label{eom00}\\
  -{\phi^2\over 6a^2}\left( 2a\ddot{a}+\dot{a}^2+k\right) &=& -{1\over 2} 
    \dot{\phi}^2 +{1\over 6}\ddot{\left(\phi^2\right)} +{1\over 3}
    {\dot{a}\over a}\dot{\left(\phi^2\right)}+ p+V_1^{VD},\label{eom11}
\end{eqnarray}
for $(\mu ,\nu)=(0,0)$ and $(1,1)$ respectively. And Eqs.~$(\ref{eom2})$,
$(\ref{conserv})$, and $(\ref{key})$ become respectively
\begin{eqnarray}
  & &\ddot{\phi}+3{\dot{a}\over a}\dot{\phi}+\left( {\ddot{a}\over a} +
   {\dot{a}^2\over a^2}+{k\over a^2}\right)\phi + V_1^{VD}{}'=0,\label{eomp}\\
  & & {dp\over dt}a^3 = {d\over dt}\left[ a^3(\rho +p)\right],\label{csv}\\
  & & {K\lambda^2\over 32\pi^2}\left(\phi^4-\left<\phi\right>^4\right)
    = 3p - \rho.\label{trace}
\end{eqnarray}

One can estimate the vacuum expectation value $\left<\phi\right>$ without 
knowing the present values of the coupling constant $\lambda $, the energy 
density $\rho_0$, and the scale factor $a_0$. 
Since the Newton's constant $G$ is related to the scalar field 
$\phi$ by
\begin{equation}
  G={3\over 4 \pi \phi^2},
\end{equation}
it follows that the rate of change of the Newton's constant, 
\begin{equation}
  {\cal G}\equiv {\dot{G}\over G} = -2{\dot{\phi}\over \phi} = 
  -{48\pi^2\rho_0 a_0^3\over K\lambda^2\phi^4 a^3}{\dot{a}\over a} = 
    {3\over 2}\left({\left<\phi\right>^4\over\phi^4}-1\right) H,
\label{Gchange}
\end{equation}
where $H\equiv \dot{a}/a$ is the Hubble parameter. The last two equalities are
obtained from Eqs.~$(\ref{csv})$ and $(\ref{trace})$ by taking $p=0$ in
the current matter-dominated universe.

The present upper bound of the rate of change of $G$ based on the time-delay
experiment is $\cite{reasen}$
\begin{equation}
  \left| {\cal G}_0\right| \leq {1\over 500} H_0.\label{G/H}
\end{equation}
The present value of $\phi$ is the Planck scale given by
\begin{equation}
\phi_0=\sqrt{3\over {4\pi G_0}}=6\times 10^{18} {\rm GeV}.
\end{equation}
Hence, 
\begin{equation}
  \left|\frac{\left<\phi\right>^4-\phi_0^4}{\phi_0^4} \right|
 = {2\over 3}{\left|{\cal G}_0\right| \over H_0} \leq {1\over 750},
\end{equation}
which means that the universe is nearly in the ground state today.

\begin{acknowledgments}
We are grateful to H. T. Cho for his introductive discussions on VD method. 
S.-Y.L. would like to thank W. F. Kao, H.-C. Kao, S.-Y. Wang and C.-W. Huang 
for useful discussions. This work was supported in part by the National
Science Council, ROC under the Grants NSC86-2112-M-001-009 and
NSC87-2112-M-001-039.
\end{acknowledgments}

\begin{appendix}

\section{Vilkovisky-DeWitt Method}

In this appendix, we would briefly introduce the Vilkovisky-DeWitt 
method$\cite{vilkov,dewitt,Kunst,HTCho}$ and calculate 
the one-loop VD effective potential for the scale invariant 
gravity $(\ref{action})$.

Let the naive metric of the space of the configuration of the quantum fields 
${\cal M}$ be $G_{ij}$, where $i$ and $j$ are indices which runs over all the 
quantum fields at every point of the whole space-time. $G_{ij}$ does not have 
to be gauge invariant. 

In the calculation of the one-loop effective potential, one needs to know the 
second derivative (variation) of the action $S[\phi ]$ with respect to the
quantum fields. 
Since $S$ is a scalar on ${\cal M}$, $S_{,i}\equiv\delta S/\delta
\phi^i$ is a vector on ${\cal M}$. If $G_{ij}$ describes a non-trivial curved 
space, one should take the covariant derivative of $S_{,i}$,
\begin{equation}
  D_i S_{,j} = {\delta\over\delta\phi^i}S_{,j}-\Gamma^k_{ij}S_{,k},
\label{D}
\end{equation}
instead of $\delta^2 S/ \delta\phi^i \delta\phi^j$ because the corresponding
effective potential should not depend on the choice of special background 
field configurations. Here the connection $\Gamma^k_{ij}$ can be written as
\begin{equation}
  \Gamma^k_{ij} = {1\over 2}G^{kl}(G_{li,j}+G_{lj,i}-G_{ij,l}),
\end{equation}
which is the Christoffel symbol.

In gauge theories, it is possible to define a gauge independent metric from the
naive one. Suppose that a general infinitesimal transformation of the fields 
is a pure gauge transformation, then
\begin{equation}
  \delta \phi^i = Q^i_\alpha \epsilon^\alpha,
\end{equation}
where $Q^i_\alpha$ is the generator of the gauge symmetry, and 
$\epsilon^\alpha$ is a parameter. In general, $\delta \phi^i$ should include
the gauge transformation part and the physical transformation part.
The line element of the general transformation $\delta \phi^i$ in ${\cal M}$ 
can be written as
\begin{equation}
  \delta s^2 = G_{ij}\delta\phi^i\delta\phi^j.
\end{equation}
Define the projection operator as
\begin{equation}
  \Pi^i_j\equiv \delta^i_j - Q^i_\alpha N^{\alpha\beta}Q^k_\beta G_{kj},
\end{equation}
satisfying
\begin{eqnarray}
  \Pi^i_j Q^j_\alpha &=& 0,\\
  \Pi^i_j \Pi^j_k &=& \Pi^i_k,
\end{eqnarray}
to project out the gauge transformation part of a general infinitesimal
transformation of the field. Here $N^{\alpha\beta}$ is the inverse of 
\begin{equation}
  N_{\alpha\beta} = G_{ij} Q^i_\alpha Q^j_\beta.
\label{N}
\end{equation}
The component of $\delta\phi^i$ in the physical space can then be defined by
\begin{equation}
  \delta_\perp \phi^i \equiv \Pi^i_j\delta\phi^j,
\end{equation}
and the line element of the physical transformation reads
\begin{equation}
  \delta_\perp s^2 = G_{ij}\delta_\perp\phi^i \delta_\perp\phi^j
  \equiv \gamma_{ij}\delta \phi^i \delta\phi^j,
\end{equation}
where 
\begin{equation}
  \gamma_{ij} = G_{ik}\Pi^k_j 
\end{equation}
is taken to be the gauge independent metric which measures the physical 
transformation part of a general transformation. Using $\gamma_{ij}$, the 
connection in the usual definition can be constructed in terms of 
$G_{ij}$ and $Q^i_\alpha$ as
\begin{equation}
  \Gamma^{(\gamma )}{}^k_{ij} = \Gamma^k_{ij} + T^k_{ij},
\label{Gam}
\end{equation}
where
\begin{equation}
  T^k_{ij} = -2Q^k_{\alpha;(i}B^\alpha_{j)} + Q^l_\sigma B^\sigma_{(i}
    B^\rho_{j)} Q^k_{\rho;l}
\label{T}
\end{equation}
with
\begin{eqnarray}
  B^\alpha_i &\equiv & N^{\alpha\beta} Q^i_\beta G_{ij},\label{B}\\
  Q^k_{\alpha;i} &\equiv & {\delta\over \delta\phi^i}Q^k_\alpha
    +\Gamma^k_{ij}Q^j_\alpha.
\end{eqnarray}
Here the convention of symmetrization,
\begin{equation}
  A_{(i} B_{j)} \equiv {1\over 2}(A_i B_j + A_j B_i),
\end{equation}
is understood.
Now the covariant derivative of $S_{,i}$ with 
the connection $\Gamma^{(\gamma )}{}^k_{ij}$ is gauge independent, so does the 
effective potential $V_1^{VD}$ constructed from it.

Now we turn to our case of the scale invariant gravity~$(\ref{action})$.
Following Vilkovisky's prescription$\cite{vilkov}$, we take the naive metric,
\begin{eqnarray}
   G_{\tilde{\phi}(x)\tilde{\phi}(y)}&=& \sqrt{-g}\delta (x-y),\label{G1}\\
   G_{g_{\mu\nu}(x) g_{\rho\sigma}(y)} &=& {1\over 2}\sqrt{-g}
   \left( g^{\mu\rho}g^{\nu\sigma}+g^{\mu\sigma}g^{\nu\rho}
     -g^{\mu\nu}g^{\rho\sigma}\right) \delta (x-y),
\label{G2}
\end{eqnarray}
and the infinitesimal transformations,
\begin{eqnarray}
  \delta\tilde{\phi} &=& -\epsilon^\alpha\partial_\alpha\tilde{\phi}
                       -\omega\tilde{\phi}
                   \equiv Q^{\tilde{\phi}}_{\epsilon^\alpha}\epsilon^\alpha 
                       +Q^{\tilde{\phi}}_\omega\omega ,\\
  \delta g_{\mu\nu} &=& -g_{\alpha\mu}\partial_\nu \epsilon^\alpha
                       -g_{\alpha\nu}\partial_\mu \epsilon^\alpha
                       -\epsilon^\alpha \partial_\alpha g_{\mu\nu} 
                       +2\omega g_{\mu\nu}
                   \equiv Q^{g_{\mu\nu}}_{\epsilon^\alpha} \epsilon^\alpha
                       +Q^{g_{\mu\nu}}_\omega \omega ,
\end{eqnarray}
where $\epsilon^\alpha$ and $\omega$ are parameters corresponding to the 
reparametrization and scale transformations respectively. Then the 
generators of these transformations are
\begin{eqnarray}
  Q^{\tilde{\phi}(z)}_{\epsilon^\alpha (x)} &=& 
    -\partial_\alpha \tilde{\phi} \delta (z-x),\\
  Q^{g_{\mu\nu(z)}}_{\epsilon^\alpha (x)} &=&
    -g_{\alpha\mu}\partial_\nu\delta (z-x)-g_{\alpha\nu}\partial_\mu\delta (z-x)
    -\delta (z-x)\partial_\alpha g_{\mu\nu}(z),\\
  Q^{\tilde{\phi}(z)}_{\omega (x)} &=& -\tilde{\phi}(z)\delta (z-x),\\
  Q^{g_{\mu\nu(z)}}_{\omega (x)} &=& 2 g_{\mu\nu}(z)\delta (z-x).
\end{eqnarray}
Here the derivative $\partial_\mu$ always acts on the first argument of 
the $\delta$ function. 
 
A straightforward calculation gives the quantities~$(\ref{N})$
evaluated at the ground-state background values~$(\ref{BF1})$
and $(\ref{BF2})$,
\begin{eqnarray}
  \left. N_{\epsilon^\alpha (x)\epsilon^\beta (y)} \right|_{bg} &=&
    -2\eta_{\alpha\beta}\partial^2\delta (x-y),\\
  \left. N_{\omega (x) \omega (y)} \right|_{bg} &=& -15\delta (x-y),\\
  \left. N_{\epsilon^\alpha (x)\omega (y)} \right|_{bg} &=&
  -\left. N_{\omega (y)\epsilon^\alpha (x)} \right|_{bg} =
  -4 \partial_\alpha \delta (x-y),
\end{eqnarray}
whose inverses read
\begin{eqnarray}
  \left. N^{\epsilon^\alpha (x)\epsilon^\beta (y)} \right|_{bg} &=&
    \left( -{1\over 2}\eta^{\alpha\beta}{1\over\partial^2} - 
    {4\over 23}{\partial^\alpha\partial^\beta \over \partial^4}\right) 
    \delta (x-y),\\
  \left. N^{\omega (x)\omega (y)} \right|_{bg} &=& -{1\over 23}\delta (x-y),\\
  \left. N^{\epsilon^\alpha (x) \omega (y)} \right|_{bg} &=&
    -\left. N^{\omega (x) \epsilon^\alpha (y)}\right|_{bg} = 
    {2\over 23}{\partial^\alpha\over \partial^2}\delta (x-y).
\end{eqnarray}
Hence, the background values of the quantities defined in Eq.~$(\ref{B})$ are
\begin{eqnarray}
  \left. B^{\epsilon^\alpha (u)}_{\tilde{\phi}(z)}\right|_{bg} &=&
    -{2\over 23}{\partial^\alpha\over \partial^2}\delta (u-z),\\
  \left. B^{\omega (u)}_{\tilde{\phi}(z)}\right|_{bg} &=& 
    {1\over 23}\delta (u-z),\\
  \left. B^{\epsilon^\alpha (u)}_{g_{\mu\nu}(z)}\right|_{bg} &=&
    \left( -\eta^{\alpha (\mu}{\partial^{\nu )}\over \partial^2} + 
    {7\over 46}\eta^{\mu\nu}{\partial^\alpha\over \partial^2} +
    {8\over 23}{\partial^\alpha \partial^\mu \partial^\nu \over \partial^4}
    \right) \delta (u-z),\\
  \left. B^{\omega (u)}_{g_{\mu\nu} (z)}\right|_{bg} &=&
    -{4\over 23}\left( {\partial^\mu \partial^\nu\over \partial^2}-
    \eta^{\mu\nu}\right) \delta (u-z).
\end{eqnarray}
It follows that $T^k_{ij}$ in Eq.~$(\ref{T})$ are given by
\begin{eqnarray}
  T^{\tilde{\phi}(z)}_{\tilde{\phi}(x)\tilde{\phi}(y)} &=&
    {-3\over 529} \delta (x-z) \delta (y-z),\\
  \eta^{\rho\sigma}T^{g_{\rho\sigma}(z)}_{\tilde{\phi}(x)\tilde{\phi}(y)}&=&
    {71\over 529} \delta (x-z) \delta (y-z),\\
  T^{\tilde{\phi}(z)}_{g_{\mu\nu}(x)\tilde{\phi}(y)} &=&
    \left( {553\over 1058}{\partial^\mu \partial^\nu\over \partial^2} +
    {277\over 1058}\eta^{\mu\nu}\right) \delta (x-z)\delta (y-z),\\
  \eta^{\rho\sigma}T^{g_{\rho\sigma}(z)}_{g_{\mu\nu}(x)\tilde{\phi}(y)}&=&
    \left( -{192\over 529}{\partial^\mu \partial^\nu \over\partial^2} +
    {146\over 529}\eta^{\mu\nu}\right) \delta (x-z)\delta (y-z),\\
  T^{\tilde{\phi}(z)}_{g_{\alpha\beta}(x)g_{\mu\nu}(y)} &=&
    \left[ -{48\over 529}{\partial^\alpha \partial^\beta \partial^\mu
           \partial^\nu \over \partial^4} 
    +{25\over 529} \left( 
      \eta^{\mu\nu}{\partial^\alpha \partial^\beta\over\partial^2} +
      \eta^{\alpha\beta}{\partial^\mu \partial^\nu\over \partial^2}\right)
    -{2\over 529}\eta^{\alpha\beta} \eta^{\mu\nu}
    \right]\delta (x-z)\delta (y-z),\\
  \eta^{\rho\sigma}T^{g_{\rho\sigma}(z)}_{g_{\alpha\beta}(x)g_{\mu\nu}(y)}&=&
    \left[ {400\over 529}{\partial^\alpha\partial^\beta\partial^\mu
           \partial^\nu \over \partial^2}
    -2{\partial^{(\mu}\eta^{\nu )(\alpha}\partial^{\beta )}\over \partial^2}
    +{313\over 529}\left(
      \eta^{\mu\nu}{\partial^\alpha \partial^\beta\over\partial^2} +
      \eta^{\alpha\beta}{\partial^\mu \partial^\nu\over \partial^2}\right)
    -{465\over 1058}\eta^{\alpha\beta} \eta^{\mu\nu}
    \right]\nonumber\\ & &\delta (x-z)\delta (y-z).
\end{eqnarray}
On the other hand, the background values of the connection of the naive metric
$(\ref{G1})$ and $(\ref{G2})$ are 
\begin{eqnarray}
  \left.\Gamma^{\tilde{\phi}(z)}_{\tilde{\phi}(x)g_{\mu\nu}(y)}\right|_{bg}&=& 
  \left.\Gamma^{\tilde{\phi}(z)}_{g_{\mu\nu}(y)\tilde{\phi}(x)}\right|_{bg} = 
    {1\over 4}\eta^{\mu\nu}\delta (x-z)\delta (y-z),\\
  \left.\Gamma^{g_{\mu\nu}(z)}_{\tilde{\phi}(x)\tilde{\phi}(y)}\right|_{bg}&=&
    {1\over 4}\eta_{\mu\nu}\delta (x-z)\delta (y-z),\\
  \left.\Gamma^{g_{\rho\sigma}(z)}_{g_{\mu\nu}(x)g_{\alpha\beta}(y)}
    \right|_{bg} &=& -{1\over 2}\delta (x-z)\delta (y-z) \left[
    \eta^{\alpha (\mu}\delta^{\nu)}_{(\rho}\delta_{\sigma )}^\beta +
    \eta^{\beta (\mu}\delta^{\nu)}_{(\rho}\delta_{\sigma )}^\alpha\right.
    \nonumber\\& & -\left.
    {1\over 2}\eta^{\alpha\beta}\delta^\mu_{(\rho}\delta_{\sigma )}^\nu -
    {1\over 2}\eta^{\mu\nu}\delta^\alpha_{(\rho}\delta_{\sigma )}^\beta -
    {1\over 2}\eta_{\rho\sigma}\eta^{\alpha (\mu}\eta^{\nu )\beta} +
    {1\over 4}\eta_{\rho\sigma}\eta^{\alpha\beta}\eta^{\mu\nu}\right].
\end{eqnarray}
Combining these with $T^k_{ij}$, the gauge independent connection~$(\ref{Gam})$
and hence the covariant derivative~$(\ref{D})$ can be worked out. This is 
equivalent to adding the correction terms,
\begin{eqnarray}
  {\cal L}'&=&{1\over 2}\sigma \left( -{12\over 529}\lambda\phi^2\right)\sigma 
  + \sigma\left( -{505\over 6348}\lambda\phi^2\right)
    {\partial^\mu\partial^\nu\over \partial^2} \sigma
  + \sigma\left( -{2\over 529}\lambda\phi^2\right) h
  + {1\over 2}h\left( {497\over 50784}\lambda\phi^2\right) h
  \nonumber\\& &
  + {1\over 2}h_{\mu\nu}\left( -{\lambda\phi^2\over 1587}\right){\partial^\mu
    \partial^\nu\partial^\alpha\partial^\beta \over\partial^4}h_{\alpha\beta}
  + {1\over 2}h_{\mu\alpha}\left( {1\over 24}\lambda\phi^2 \right)
    {\partial^\alpha\partial_\beta\over\partial^2}h^{\mu\beta}
  + h\left( -{171\over 8464}\lambda\phi^2 \right)
    {\partial^\mu\partial^\nu\over\partial^2} h_{\mu\nu},\label{L'}
\end{eqnarray}
to the original quadratic Lagrangian~$(\ref{Lq})$. 

\end{appendix}

\end{document}